\documentclass[referee]{aa}
\usepackage{graphics}

\def\beq{\begin{equation}}
\def\eeq{\end{equation}}
\def\bey{\begin{eqnarray}}
\def\eey{\end{eqnarray}}
\def\kms{\rm \,km\,s^{-1}}

\def\rv{\Delta{v}}
\def\SN{S\!/\!N}

\def\corr#1#2{\langle{#1,#2}\rangle}

\begin{document}

\thesaurus{04         
	(03.20.8;  
	03.13.2;   
	12.07.1;   
	08.11.1)}  

\title{Detecting Luminous Gravitational Microlenses Using Spectroscopy}

\author{S. Mao \inst{1} \and
	J. Reetz \inst{2} \and D.J. Lennon \inst{2}}

\offprints{S. Mao}
\mail{smao@mpa-garching.mpg.de}

\institute{Max-Planck-Institut f\"ur Astrophysik, 
	Karl-Schwarzschild-Strasse 1, 85740 Garching, Germany
\and Universit\"ats-Sternwarte, Scheinerstrasse 1, 81679 Munich, Germany}

\date{Received 1998; accepted 1998}
\titlerunning{
Detecting Luminous Gravitational Microlenses Using Spectroscopy}
\authorrunning{Mao et al.}

\maketitle

\begin{abstract}

We propose a new method to detect the gravitational lenses in the
ongoing microlensing experiments using medium and high resolution spectroscopy
($\lambda/\Delta \lambda > 6000$).
Since the radial velocity of the lens and lensed source typically differs
by $\sim 100\kms$, the spectral lines from the lens and source will be shifted
relative to each other by $(1-2)$\AA\ in the optical.
We simulate realistic composite spectra assuming different spectral types
for the lens and source and study the lens detectability as a function of
the signal-to-noise ratio, spectral resolution
and lens-to-source light ratio. We show that
it is possible to measure the difference in radial velocity
from an unequivocal signature in the difference of
cross- and auto-correlation functions calculated from
two spectra obtained at different magnifications.
If the lens is brighter than 10\% ($\Delta{\rm m}_{\rm v}\sim 2.5$)
of the unmagnified source
we find that a spectral resolution of $\lambda/\Delta \lambda \sim 6000$
and a signal-to-noise of 50 (at magnification maximum)
are sufficient to determine the relative radial velocity of the lens.
At $\lambda/\Delta\lambda = 40000$, the spectral resolution of high
resolution spectrographs of 8-10m class telescopes,
the lens could even be detected at a brightness of $\sim 3\%$ 
($\Delta{\rm m}_{\rm v}\sim 4.0$) of the source.
Radial velocities higher than $50\kms$ can be measured with an accuracy
of a few $\kms$.
Practical difficulties and
observation strategies are also discussed.

\keywords{
Gravitational Lensing --
Stars: kinematics --
Techniques: spectroscopic --
Methods: data analysis
}

\end{abstract}

\section{Introduction}

Gravitational microlensing has been rapidly maturing in the last few years
(see Paczy\'nski 1996 for a review).
Up to now, about two hundred microlensing events have been detected,
about 15 of these are toward
the Large and Small Magellanic Clouds (LMC, SMC,
Alcock et al. 1993, 1997b; Aubourg et
al. 1993) while the rest 
are toward the Galactic bulge (Alcock et al. 1995, 1997a;
Udalski et al. 1994a, 1994b; Alard \& Guibert 1997).
Thanks to the early warning/alert systems (Udalski et al. 1994c;
Alcock et al. 1996),
microlensing events are now routinely identified in real-time.
This allows simultaneous photometric (Alcock et al. 1997c; Albrow et
al. in preparation)
and spectroscopic observations (Lennon et al. 1996, 1997).
To date, for all the microlensing events, the intervening lens has
never been convincingly detected toward either the Galactic bulge or
the LMC. 
Since the lens and source are aligned to about
$10^{-3}$ arcsecond, if the lens is luminous, 
the observed light curve should include the light
from the lens as well as the source. 
In fact towards the Galactic bulge the lenses are expected to be normal
stars. The situation is less clear toward the LMC, however if some of the
lenses are within the LMC itself (Sahu 1994; Wu 1994) or in tidal
debris or streams (Zhao 1998), 
then these should be luminous as well.
As it is most likely that the lens is less massive and
considerably fainter than the source, the problem then is how to
detect its presence.  Photometric methods rely on the fact that
the lens contribution will distort the light curve and potentially 
this can be used to detect the lens (Di Stefano \& Esin
1995; Kamionkowski 1995; Buchalter et al. 1996).
So far this photometric method has not yielded
a reliable detection of the lens due to the source blending problem in the
crowded fields (see discussion).
In this paper, we suggest a spectroscopic method
to infer the presence of the lens.
This method is based on the fact
that the lens and the source have radial
velocities that differ by $\sim 100\kms$,
and hence the observed spectra should include stellar lines
from both components shifted relative to each other by
$(1-2)$\AA\ in the optical range;
the required resolution to detect such a line shift is therefore not
exceedingly high. The detectability of the lens component in the
composite is obviously a function of the ratio of the 
brightnesses of the source and the lens, and their intrinsic
spectral distributions.  
The situation is analogous to the detection of spectroscopic
binaries, except that  there are two essential differences:
the lens is normally expected to be much fainter than the source and
the light ratio changes during the microlensing event.  We make
use of this latter effect in the method presented here, and,
using simulated composite spectra, we have
experimented with different methods to detect the lens.
In brief we find that the differential cross-correlation of two
spectra obtained at different epochs offers an efficient way of
detecting the lens;
this method is explained in Sect. 2 and the simulation results
are presented in Sect. 3. We outline practical observing strategies and
discuss our results in Sect. 4.

\section{Method}

A common method to measure radial velocities is the
cross-correlation technique. The cross-correlation function for two
functions $Y(x)$ and $h(x)$ is defined 
as 
\beq
\corr{Y}{h}(\xi) =
      \frac{1}{x_2 - x_1}\int\limits_{x_1}^{x_2} Y(x) h(x+\xi)\,dx,
\eeq
where $x$ is a function of wavelength.
This method works well if the spectra $Y$ and $h$ focus their spectral 
power in similar features. Let $h$ be a single component radial velocity 
standard. If $Y$ is a single component spectrum, the radial velocity
simply corresponds to the position of the correlation maximum.
For a two-component spectrum, such as in our case, 
one might expect that the relative radial velocity can be measured
from the distance of the correlation maxima, provided 
these are well separated and distinguishable from sidelobes. 
However, most lenses are expected to be much
fainter than the source, and moreover the source itself is
faint (with visual magnitudes typically between 19 and 21 for the
Galactic bulge). In this case the use of a direct cross correlation technique
is inappropriate, especially when the spectral resolution is low,
since it becomes difficult to detect the correlation maximum
due to the lens (see Fig. 1). 
We therefore experimented with other estimators. 
We found that a differential cross-correlation
of composite spectra observed at two different magnifications
(Eq. \ref{equ:diff-corr}) provides a more sensitive signature
to determine the relative radial velocity 
between the lens and source.

Suppose we have a series of spectra obtained during a microlensing event. 
Each spectrum is a composite of the lens and (amplified) source:
\beq \label{equ:composite}
Y_i(x) = \mu_i\cdot S(x) + L(x) + N_i(x), ~~ x \equiv \ln\lambda,
\eeq
where $i$ is a sequence number, $S(x), L(x)$ are the intrinsic source and
lens spectrum respectively, $N_i(x)$ is the noise term, and $\mu_i$ is the
(intrinsic) amplification of the source.
The observed spectra are sampled logarithmically, so a velocity shift
is a uniform linear shift in $x$.
Radial velocity measurements are usually performed using
rectified spectra, i.e., spectra which are normalized to the continuum,
where the continuum is determined by an interpolation of the
observed spectrum by low-order polynomials. From here on, we will assume that 
$Y_i$ has been rectified using standard procedures. Since
we determine the radial velocity by comparing the relative shift in the
positions of spectral lines, we are only interested in the alternating
part of $Y_i(x)$ coming from the lines.
Therefore we subtract the mean of $Y_i(x)$
and normalize the spectrum by its {\it rms}:
\beq 
y_i(x) = {Y_i(x) - \overline{Y_i} \over \sigma_{i}},
\eeq
where $\overline{Y_i}$ and $\sigma_{i}$ are the mean and rms
of the rectified spectrum within a certain
spectral range $x_1 < x < x_2$. 
Note that this normalization, which brings each composite spectrum onto
the same scale, requires neither an accurate
knowledge of the (intrinsic) magnification nor that of
the absolute flux distribution. This property is one of the benefits
compared to a method which makes direct use of Eq. (\ref{equ:composite})
to disentangle $S$ and $L$ algebraically from at least two 
different observations $Y_1$ and $Y_2$.

Spectral regions with many well separated intrinsic
atomic and molecular lines are well suited for 
correlation studies.
Non-intrinsic features such as interstellar Na\,D lines, telluric
or night sky lines should be avoided if the residues of these lines 
have significant spectral power.
In practice, we choose the spectral window between 5100\,\AA\
and 5700\,\AA\ which includes mostly narrow metal lines.
At low resolution, the spectral power is dominated by
strong well separated features such as Mg\,b triplet,
and few strong {\rm Fe}{\,\sc i} lines.
We avoid the Balmer lines (e.g., H$\beta$) for the correlation
because the significance of the radial velocity indicator
varies with the spectral types if these temperature sensitive
lines are included.

The differential correlation function of two spectra, $y_1$ and $y_2$,
is defined as
\beq
 \label{equ:diff-corr}
 \corr{{y}_1}{\Delta {y}} =
    \corr{{y}_1}{{y}_2} - \corr{{y}_1}{{y}_1},
\eeq
where $y_1$ should be taken as the spectrum with the highest
$\SN$, e.g., a spectrum observed near the maximum of
magnification. The second term (auto-correlation of $y_1$)
in Eq. (\ref{equ:diff-corr}) is symmetric, but the first term
(cross-correlation term) is in general asymmetric, so the differential
correlation function is asymmetric. It is easy to show that when the
amplification is very high for $y_1$, the differential correlation function is
essentially reduced to the cross-correlation of
the source and the lens spectrum. The maximum of the
differential correlation function simply corresponds to the 
difference in radial velocity between the lens and the source. 
We evaluate the errors in the relative radial velocity 
by comparing the derived radial
velocity in simulations with different noise patterns.

\section{Simulations}

The composite spectra were created using rectified synthetic
lens spectra, $l(x)$, and source spectra, $s(x)$,
\beq
Y_i(x) = \frac{\mu_i\cdot s(x) + \gamma_0 \cdot l(x)}{\mu_i+\gamma_0},
\eeq
where $\gamma_0$ is the lens-to-source light ratio when the source is
at the baseline, i.e., when it is unamplified.
We investigate three combinations of the lens and source
spectra: a K0 star lensed by a K5 star, an F5 star lensed by a
K5 star and an F5 source by a solar-type (G2) lens.
The choice of source spectral types is guided by the fact
that most sources identified in the bulge have magnitudes and
colours implying that they are main sequence or turn-off
stars of spectral types F to K.  This is supported by
recent spectroscopic observations
(Lennon et al. 1996, 1997).
The choice of the lens spectral types is more
problematic since we cannot at present generate
convincing M-dwarf spectra.  We therefore restrict
ourselves to the cases of K and G dwarfs, thereby limiting
the lens masses ($m$) which 
we sample to $m > 0.5{\rm M}_\odot$.
We return to this very important point in the discussion.
Theoretical spectra are calculated using an extended grid of
homogeneous, plane-parallel model atmospheres described by
Fuhrmann et al. (1997).
LTE-line formation was performed with molecular and atomic lines from
Kurucz (1992). The $f$-values for most lines within
$\lambda\lambda = 4800 - 5500$\AA\ have been adjusted to the
solar flux atlas (Kurucz et al. 1984).
We find that the significance of the
cross-correlation signatures depends on the assumed combination of
spectral types (and spectral correlation range).
However the physical approximations
of the models and the accuracy of the line data
have little impact on the results.
For each spectrum, the spectral resolution has been
reduced to the desired value by a convolution with Gaussian profiles;
then normally distributed noise pattern are added according to the
spectral $\SN$.
The $\SN$ of the composite when amplified by $\mu_i$ is given by
$
  (\SN)_{(i)}   = (\SN)_0
                (\frac{\mu_i+\gamma_0}{1+\gamma_0})^{1/2},
$
where we have assumed that the $\SN$ is dominated by photon noise.
The baseline signal-to-noise ratio, $(\SN)_0$, has been scaled from
our previous real-time spectroscopic observations of microlensing events
(Lennon et al. 1996, 1997) or from technical specifications of typical
high-resolution spectrographs for 8-10m class telescopes. 
The difference in radial
velocity of the lens and source is taken to be $50-100\kms$.
This range is chosen because of the following reasons:
Toward the LMC, the random radial motion of
halo lenses is expected to be $V_c/\sqrt{2}=160\kms$, where $V_c \approx
220\kms$ is the circular velocity of the halo.
If the LMC self-lensing
contributes significantly to the optical depth, then the LMC lenses are
expected to be kinematically hot, with velocity dispersion
$\sim 60\kms$ (Gould 1995).
The lenses and sources in
the Galactic bulge have velocity dispersions of about 100$\kms$. Therefore
the difference in radial velocity between the
lens and source is about $100\kms$.
Since the lens is expected to be much fainter than the source,
we have studied cases where
the lens-to-source light ratio at the baseline
is between 0.1 to 0.01, corresponding to differences
in apparent magnitude ($\Delta{\rm m}_{\rm v}$) of 2.5 and 5.0.

We first illustrate the difficulty in detecting the lenses using the
standard cross-correlation technique. In Fig. 1, we show the
auto-correlation functions for $\gamma_0=1/25, 1/75$ for two
spectral resolutions, $\lambda/\Delta\lambda=6000, 40000$
with $\rv=100\kms$ and $50\kms$, respectively. One
clearly sees that there is no secondary correlation maximum associated
with the lens: It is simply lost in side lobes.
\begin{figure*}
\resizebox{\hsize}{!}{\includegraphics{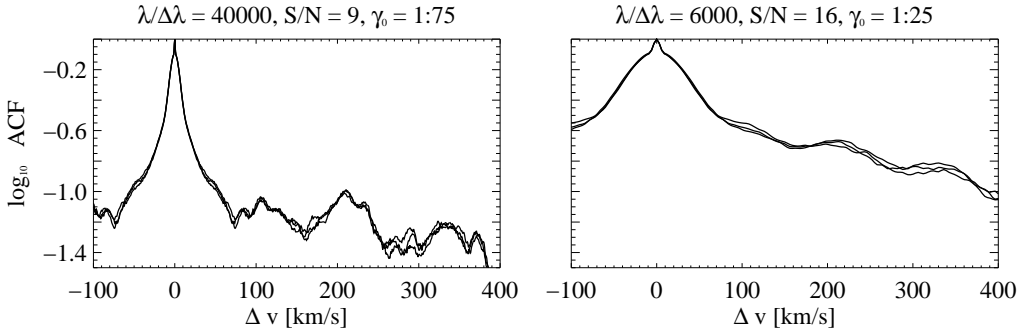}}
\caption{
Auto-correlation functions in logarithm are shown
for two composite spectra where the baseline
lens-to-source light ratio, $\gamma_0$,
is 1/75, 1/25 respectively. 
The spectral resolution is taken to be 
40000 and 6000 respectively.
The lens and source have a difference
in radial velocity of $\Delta v = 50\kms, 100\kms$ for these two resolutions.
The adopted $\SN$ is expected
for a $V \sim 20$ star with one hour exposure for an 8-10m class
telescope.
For each case, three curves with
different noise patterns are shown. Notice there is {\it no} secondary
peak at $\Delta v = 50\kms$, $100\kms$.
}
\end{figure*}

Since the minimum spectral resolution to detect the line shift is given
by $\lambda/\Delta\lambda=c/\Delta V=3000$, we first study simulations
with medium resolution
$\lambda/\Delta\lambda=6000$. The $\SN$ is chosen to be between 10 and 16
at the baseline and the $\SN$ at the peak amplification is taken to be
50, implying magnifications of 25 and 10 respectively; such peak
magnifications are typical for the observations conducted by Lennon et
al. (1996, 1997).
Fig. 2 shows the results for $\gamma_0=0.1$ and 0.04,
respectively. It appears that when the lens-to-source light ratio is
0.1, the radial velocity can be inferred with an accuracy of $\la
10\kms$. This determination
is least reliable when the lens is a G2
star and the source F5, because the density of
spectral lines from molecules and neutral metals
is lower at higher effective temperatures.
When the lens-to-source light
ratio is reduced to 0.04, the reliability is still acceptable when the lens
is a K5 star, but it becomes worse when the source is an F5 star
and the lens a G2 star: two of the three simulations show large
departures from the input
radial velocity, indicating for such low values of $\gamma_0$, the
differential correlation method is not sensitive enough.
It appears that with spectral resolution of about
6000, the lens can be detected reliably when the intrinsic (unamplified)
lens-to-source light
ratio is about $\sim 10\%$. 
Note that the significance of the correlation maximum is better
for the case where the lens and source are both cool stars.
This is mainly caused by an increase in the number of 
metal lines as the effective
temperature decreases. 

\begin{figure*}
\resizebox{\hsize}{!}{\includegraphics{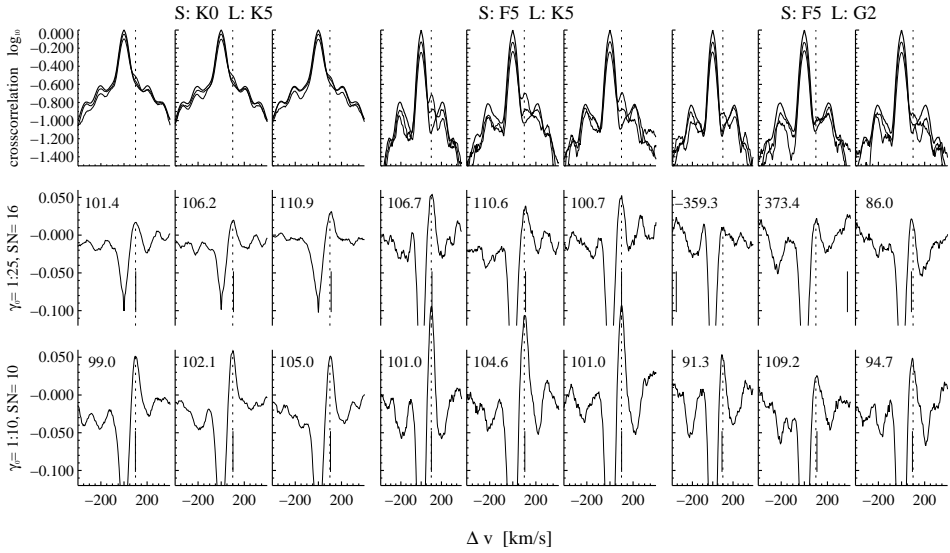}}
\caption{ \label{fig:lores}
Differential correlation functions (Eq. \ref{equ:diff-corr})
are shown for three lens (L) and source (S) spectral
combinations (indicated at the top). 
The spectral resolution is $\lambda/\Delta\lambda=6000$, and the
difference in radial velocity between the lens and source is $100\kms$.
The baseline $\SN$ is 10 and 16, 
and the lens-to-source light ratio is 0.1 and 0.04, respectively.
The peak $\SN$ is taken to be 50, implying magnifications of
25 and 10 respectively.
For each combination of the lens and source spectral types
simulation results are shown for three 
noise patterns, along with the estimated
$\rv$ indicated in each sub-panel.   
The corresponding auto-correlation ($\corr{y_1}{y_1}$) and two
cross-correlation ($\corr{y_1}{y_2}$) functions are shown
in the first row, here $y_1$ is the peak spectrum, and $y_2$ is the
two baseline spectra with $\SN=10$ and 16.
Notice that their differences are very small.
The dotted line marks the input $\rv$ whereas
the solid vertical line indicates the detected correlation maximum.
}
\end{figure*}

\begin{figure*}
\resizebox{\hsize}{!}{\includegraphics{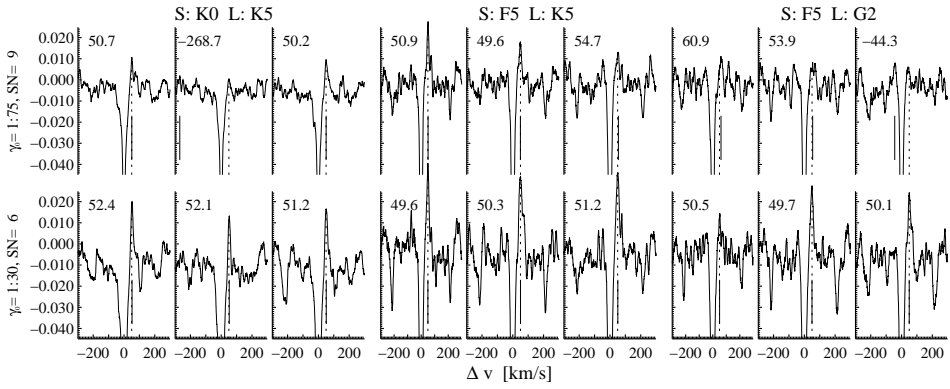}}
\caption{
\label{fig:hires}
Differential correlation functions (Eq. \ref{equ:diff-corr})
are shown for three lens and source spectral
combinations (labelled at the top). 
The input $\Delta v$
is $50\kms$ and
the spectral resolution is
$\lambda/\Delta\lambda=40000$, appropriate for the UVES instrument on VLT.
The symbols are the same as in Fig. 2. The baseline $\SN$ 
is 6 and 9, and the lens-to-source light ratio is 3.3\% and 1.3\%
respectively. The peak $\SN$ is 30, implying magnifications of 25 and
10, respectively.
}
\end{figure*}

We next consider the situation for 
high-resolution spectrographs available on 8-10m
class telescopes. We will take the UVES instrument to be installed on
the VLT (Pilachowski et al. 1995) as an example.
The $(\SN)_0$ is taken to be 6 or 9 when the
source is unmagnified;
this corresponds to a $\SN$ achieved for a $V \sim 20$
star with a one-hour exposure. Fig. \ref{fig:hires} shows numerical
results for $\rv = 50\kms$ for a lens-to-source light ratios of 1.3\%
and 3.3\%. The $\SN$ at the maximum is 
taken to be 30, implying magnifications of 30 and 10 respectively.
For the case of $\gamma_0=1.3\%$, some inferred radial velocity
show large deviations from the input value. For $\gamma_0=3.3\%$, the
situation is much improved. The
radial velocity is recovered reliably, 
with an accuracy of $\la 2 \kms$, 
for all combinations of lens and source spectral distributions. 
We have ran simulations with
radial velocity differences as low as $20\kms$
for $\gamma_0=3.3\%$,
and the radial velocity
difference can still be inferred quite reliably.
It is clear that with such high resolutions, the stellar lines
are resolved, although contaminated by noise; 
the differential cross-correlation yields very
reliable radial velocities for a lens-to-source light ratio of $\ga 3\%$.
Therefore, the detection of microlenses with such high-resolutions
is clearly feasible.

So far we have mainly concentrated on the high-magnification
microlensing events that
allow us to obtain high S/N ratio spectrum at the peak, however,
events with low magnifications are more common (the lensing frequency roughly
scales as $A^{-1}$), we have therefore also ran simulations for a case
with a peak magnification of 2.5. We found that in such
cases, we can still detect the lenses spectroscopically when 
$\gamma_0 \ga 10\%$ with baseline
S/N $\ga  10$. It appears that our method
works rather well for even modestly magnified microlensing
events. However, in reality, 
highly magnified microlensing events are still preferred since the high
S/N spectrum 
at the peak allows the intrinsic source properties be more realiably
derived; such properties are valuable for not only lensing studies but
also for the kinematical and chemical evolution of the Galactic bulge
(Lennon et al. 1997).

\section{Summary and Discussion}

As we have implied above, the choice of our spectral types
does not sample lens masses below about 0.5 solar masses.
If we assume that the lensing in the bulge is primarily self-lensing,
such that lens and source both reside there, then this implies
a mean lens mass of around 0.3-0.5 solar masses (depending on the mass
function).
If the most likely lens is an M dwarf ($\sim 0.3 M_\odot$) with an absolute
visual magnitude of around M$_V\sim +10$ (compared with
M$_V\sim +7$ for a K5 dwarf), while a typical turn-off star in the
bulge would have M$_V\sim +3$ which implies that the
light ratio would be of order 10$^3$. 
Detecting such a lens is
clearly much more difficult and therefore it is 
necessary to investigate the fraction of lens with masses above about 0.5 
solar masses. The luminosity function of the bulge stars implies  
a mass function of $\phi(m) dm \propto m^{-2.2} dm$ for 
$0.7M_\odot<m<1 M_\odot$ and
$\phi(m) dm \propto m^{-1} dm$ for $0.3 M_\odot < m<0.7 M_\odot$
(Holtzman et al. 1998),
since the event rate scales as $m^{1/2}$,
one can estimate that roughly 45\% of the lenses have mass above $0.5
M_\odot$; in the above estimate, we have ignored the detection efficiency
as a function of event duration since each lens mass produces very broad
event duration due to the convolution of kinematics and lens distances
(e.g., Mao \& Paczy\'nski 1996).
This estimate is still uncertain, since
the mass function of the bulge implied by the microlensing experiments
is still under debate and of course is an important
objective of microlensing surveys
(see, e.g., Han \& Lee 1997; Zhao \& de Zeeuw 1997; see Gould 1996 for a
review). In this respect
the detection of even only the most luminous lenses is obviously
very important.

We have shown through realistic simulations that it is possible
to detect the presence of the lens using high-resolution spectrographs
available on 8-10m class telescopes (Pilachowski et al. 1995),
even when the lens to source light ratio is as small as $\sim 3\%$.
This method complements the photometric detection of blending in
microlensing. The spectroscopic observations will yield further
kinematical information on the lens and source, which will provide
more constraints on the modelling of microlensing events.
The observations will require flexible scheduling of telescope time. Such
real-time spectroscopic
observations have already been carried out successfully by Lennon et
al. (1996, 1997) using the NTT at ESO. Unfortunately,
the spectral resolution used, $\lambda/\Delta\lambda \approx 1200$, was
too low to detect the relative radial velocity as our numerical
simulations suggest.
The proposed spectroscopic observations should first be performed
with the bulge microlensing events for two reasons. First,
the number
of events is a factor of ten larger toward the bulge, therefore
we can select highly amplified events to perform high $\SN$ observations.
Second, the bulge sources are brighter than those in LMC.
Notice that the lens-to-source light ratio obviously increases
when the source becomes fainter. The best cases to study the
lenses are therefore highly amplified faint stars.
If the proposed method works for
bulge microlensing events, then it is clearly very interesting to extend
such high resolution observations toward the LMC.

Detailed photometric follow-up observations make
the selection of spectroscopic observations simpler since the
photometric effect detected in real-time can already tell us whether
some additional light from the lens or a nearby source is present by chance
alignment. The second possibility arises because
the bulge and LMC are both crowded; many
of the current observed microlensing events are influenced by such
blending, since 
they exhibit astrometric centroid shifts of a few tenths of arcseconds
during microlensing (Goldberg \& Wozniak 1997). Photometric
methods to detect the lenses (e.g., Kamionkowski 1995)
are unable to tell whether the
additional light is from the lens or from a blended source. Our method
is subject to the same difficulty, although here the derived kinematical
information will be helpful since the transverse motions (and hence 
indirectly the radial velocities) is subject to the duration constraint
while that of a random blending star is not.
The proposed method can of course be used to confirm such
blending. However if one is more interested in detecting the lens,
events with astrometric shifts should be eliminated.
The seeing expected on the sites of 8-10m class telescopes will be
substantially better than the typical seeing 
at the current microlensing survey sites ($\ga 1.5^{''}$).
Many of the blended events will be spatially resolved. 
Furthermore, with the superior resolution of HST, one can largely
tell whether the light is from the lens or a random star along the line
of sight. If the additional light is not from a
random star, then it is either from the lens or
from a close binary. The latter possibility can be detected either
photometrically from the light curve (Han \& Gould 1996)
or spectroscopically from the
velocity shift in different epochs. The combination of HST and
spectroscopic observations on 8-10m class telescopes therefore provides
a powerful way of probing the lenses.

In this work, we have tried to address the
rather modest goal of detecting the lens using spectrocopic techniques.
The presence of the lens is inferred from the peak position of
differential correlation function. It is clear that the peak height is
correlated (in some way) with the lens-to-source flux ratio
(see Figs. 2 and 3). However, it
remains to be seen whether this quantity can be reliably deteremined
when the lens spectral type is unknown and with realistic S/N.
Clearly it will be very exciting if the lens-to-source flux ratio can be
inferred, since this will provide additional constraints on the lensing
parameters (Gould \& Loeb 1992; Gaudi \& Gould 1997). Although the
proposed differential correlation technique seems to work reasonably
well, it will be very interesting to explore other statistical
techniques such as the TODCOR method developed by Zucker \& Mazeh (1994)
in the context of spectroscopic binaries. We plan to address
these important issues in a further work.

\begin{acknowledgements}

We are grateful to Thomas Gehren, Peter Schneider, Klaus Simon
and Hongsheng Zhao for useful discussions. Insightful comments
by the referee, Andrew Gould, have much improved the paper.
We also thank Penny Sackett and Bohdan Paczy\'nski for encouragements.
This project is partly supported by
the ``Sonderforschungsbereich 375-95 f\"ur Astro-Teilchenphysik'' der
Deutschen Forschungsgemeinschaft.

\end{acknowledgements}


\end{document}